# Composite THz materials using aligned metallic and semiconductor microwires, experiments and interpretation

**Anna Mazhorova,[1] Jian Feng Gu,[1] Alexandre Dupuis,[1]**

**Ozaki Tsuneyuki[2], Marco Paccianti[2], Roberto Morandotti[2], Hiroaki Minamide,[3] Ming Tang,[3] Yuye Wang,[3] Hiromasa Ito[3]**

**and Maksim Skorobogatiy[1,*]**

[1] *Ecole Polytechnique de Montreal, Génie Physique, Québec, Canada*
[2] *INRS, Varennes, Québec, Canada*
[3] *RIKEN, Sendai, Japan*
[*] maksim.skorobogatiy@polymtl.ca

**Abstract:** We report fabrication method and THz characterization of composite films containing either aligned metallic (tin alloy) microwires or chalcogenide $As_2Se_3$ microwires. The microwire arrays are made by stack-and-draw fiber fabrication technique using multi-step co-drawing of low-melting-temperature metals or semiconductor glasses together with polymers. Fibers are then stacked together and pressed into composite films. Transmission through metamaterial films is studied in the whole THz range (0.1-20 THz) using a combination of FTIR and TDS. Metal containing metamaterials are found to have strong polarizing properties, while semiconductor containing materials are polarization independent and could have a designable high refractive index. Using the transfer matrix theory, we show how to retrieve the complex polarization dependent refractive index of the composite films. We then detail the selfconsistent algorithm for retrieving the optical properties of the metal alloy used in the fabrication of the metamaterial layers by using an effective medium approximation. Finally, we study challenges in fabrication of metamaterials with sub-micrometer metallic wires by repeated stack-and-draw process by comparing samples made using 2, 3 and 4 consecutive drawings. When using metallic alloys we observe phase separation effects and nano-grids formation on small metallic wires.

## References and links

## 1. Introduction

Passive optical devices in THz region, such as beam polarizers, polarization compensators, filters, etc. have attracted considerable interest owing to their numerous potential applications in THz imaging and spectroscopy [1,2]. Due to a relatively large size of optical wavelength in THz regime, there is a strong interest in realizing THz optical devices using microstructured materials operating in the effective medium regime (metamaterial regime). Particularly, if the size of the features making a composite is much smaller than the wavelength of light, then such a composite can be thought of as a homogeneous (however, anisotropic) medium [3,4] that can exhibit various unusual properties. One possible implementation of a THz metamaterial is in the form of the periodic metallic split-ring resonator arrays [5,6] which are typically fabricated by electron-beam lithography or photolithography, on a substrate made of a glass or a semiconductor wafer [7]. Another popular composite material, which is also the subject of this paper, is an array of subwavelength metal wires embedded into uniform dielectric material [8]. At THz frequencies such composites have already been successfully exploited to create polarizers and high-pass filters [9]. For example, free standing low-loss wire-grid polarizers behave as low-loss optical elements with high polarization extinction ratio. Fabrication of the terahertz wire-grid polarizers with a uniform period and fine pitch by using winding metallic wires is, however, technologically challenging, thus resulting in high cost for the large area polarizer components. The inkjet printing method is the simplest method that can be used to fabricate wire-grid polarizers [10,11]. Main limitation of such a fabrication process is the limited resolution of the printer head. Wire arrays have also been demonstrated by electrodeposition into porous alumina templates or by drilling techniques [12,13]. While proof-of-concept metamaterials were successfully demonstrated, using these fabrication techniques, however, their potential for the large scale production of THz metamaterials is still under investigation.

In this paper we describe a novel method for the fabrication of THz metamaterial films via fiber drawing and subsequent pressing of the metallic or dielectric microwire arrays into composite films. With this technique we demonstrate metal wire based metamaterial films having strong polarization properties, and chalcogenide-based metamaterial films with designable high refractive index. We would like to note that mass production of the vertically aligned extremely long metallic micro/nanowires using fiber drawing technique was recently detailed in [14]. In application to THz wire-grid polarizers there were two recent reports where the authors suggested the use of several aligned fibers containing wire arrays to build large area polarizers [15,16]. For applications in the visible spectral range there were several recent reports of micro- and nanowire fabrication by co-drawing copper or gold with glasses [17,18]. However, all these works mainly focused on the fabrication and optical properties of the individual fibers containing metallic wires, and no attempt was made of fabricating a continuous planar sample or a metamaterial film. Also, fabrication of the high refractive index composites made of chalcogenide wire arrays was not reported before. Additionally, no study has been reported on the quality of micro-wire structure after a large number of consecutive drawings. Finally, although characterization of the dielectric tensor of microwire arrays was attempted in earlier works [19], no rigorous effort was made to compare experimentally obtained metamaterial parameters with predictions of a theoretical model.

In this paper, we report a densification technique where individual metamaterial fibers are first aligned and then pressed at elevated temperature to form a continuous uniform metamaterial film. Transmission measurements through such films are then reported and a transfer matrix-based analytical model of transmission trough a 3-layer metamaterial film is then used to extract the complex frequency dependent effective refractive index and permittivity of a composite medium. Various practical challenges are detailed when trying to infer the metamaterial parameters from the transmission data. Finally, we demonstrate strong polarization effects of the metal containing metamaterials, as well as ability to design high refractive index composite materials using chalcogenide nanowires.

## 2. Fabrication process

The microwire arrays are made by the stack-and-draw fiber fabrication technique using co-drawing of low melting temperature metals or amorphous semiconductor glasses together with polymers. The fabrication process of both metal and chalcogenide glass ($As_2Se_2$) samples is similar and can be divided into several steps. First, we prepare microstructured fibers containing microwire arrays. For that we start by filling a single PC (Polycarbonate) tube with a liquid melt ($T_m$ ~ 130˚C) of Bismuth and Tin alloy (42% Bi, 58% Sn). After the first perform is made it is drawn at 190˚C into a polymer-clad wire with diameter of ~ 260 µm (first drawing). We then stack hundreds of wires together, clad them with another PC tube and draw it again under the same temperature. A scanning-electron-micrograph of the cross-section of a composite fiber after the second drawing is shown in Fig. 1(a). Inset is a magnification of an individual wire of size ~10 µm. Note that after the first drawing, metallic wire is surrounded by a polymer cladding of thickness comparable to the size of a wire. Relatively thick cladding is needed to preventing the individual wires from coalescing with each other during the following consecutive drawings.

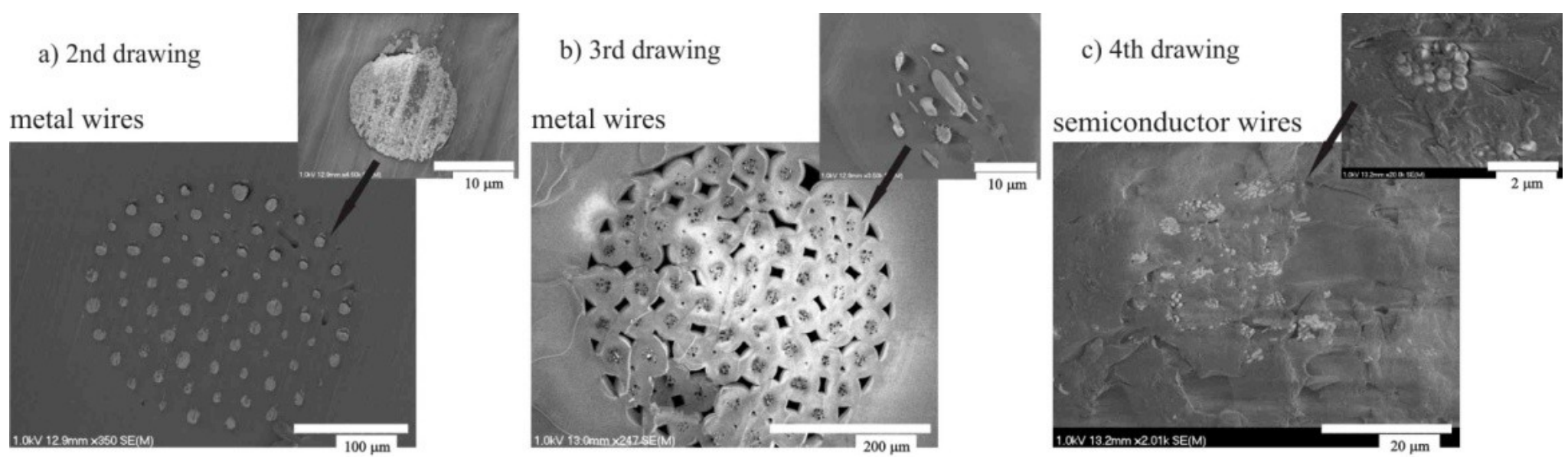

Fig. 1. SEM pictures of the cross-sections of fabricated wire-array metamaterial fibers. a) metal wire fiber after 2nd drawing, b) metal wire fiber after 3rd drawing, c) semiconductor wire fiber after 4th drawing. Insets show magnification of individual wires. Inset of c) shows cluster of nanowires with the individual fiber diameters as small as 200nm.

By repeating stack and draw process several times we can create ordered wire arrays with sub-micron size wires. The inset in Fig.1 (c) shows crossection of a fiber after the $4^{th}$ drawing featuring semiconductor wires with diameters as small as 200 nm. Finally, to create metamaterial films we use fibers after the $3^{rd}$ drawing, place them on a flat surface next to each other and then press them under pressure of several tons at 195˚C. Fig. 2 presents optical micrographs of the cross-sections of a film containing metal (Fig. 2(a)) and semiconductor (Fig. 2(b)) microwire arrays. When pressing the fibers a three-layer film is created. The fiber polymer cladding creates the two outer polymer layers that sandwich a metamaterial layer.

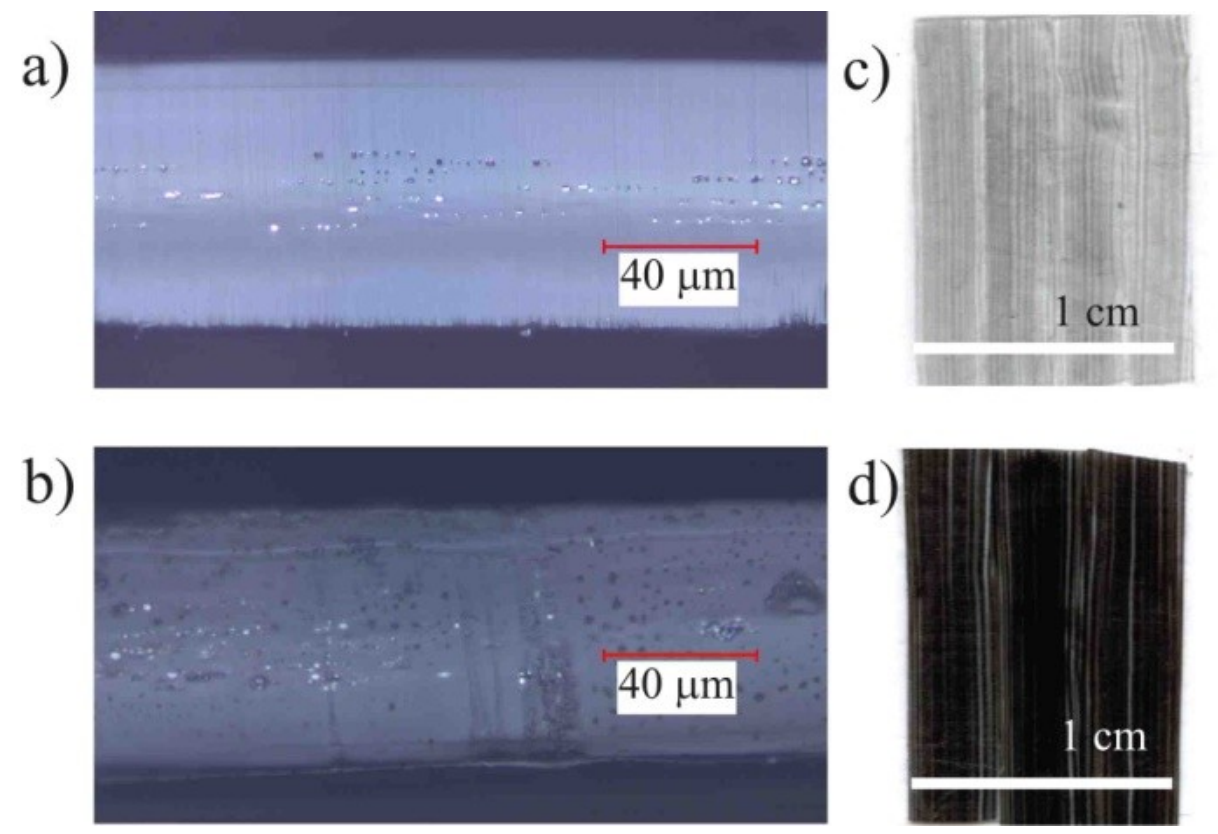

Fig. 2. Optical micrographs of a film containing (a) metal, and (b) semiconductor microwire array. Metamaterial layer is sandwiched between the two polymer layers. Figures (c) and (d) present top view of the films where metal and semiconductor wires can be seen to traverse the entire span of the film.

Fig.2 (c,d) presents an optical micrograph displaying the top view of a metamaterial film, where the aligned metal (c) or chalcogenide (d) wires can be clearly seen. From the SEM images of the pressed films we see that the wire size fluctuates in the range 0.9–4 μm. To create semiconductor nanowires we follow the same fabrication procedure but start with $As_2Se_3$ semiconductor chalcogenide glass rods cladded with a PSU (PolySulphone). Fibers were then drawn at 300˚C and subsequently arranged and pressed into films at temperatures 280˚C and pressures of several thousands of PSI.

A single fiber drawing yields hundreds of meters of a uniform diameter fiber. Cutting, stacking, and pressing such fibers into films enables fabrication of large area composite films. In our experiments we have produced films having an area of several $cm^2$, with the main limitation being the size of a hot press. The main challenge in pressing the composite films is to guarantee the uniformity of a metamaterial layer. Because of the relatively thick plastic cladding of the individual wires, most of the pressed films show stripes of metal wires separated by the stripes of clear plastic. Press parameters have to be chosen carefully to guarantee that during the compactification process the wires are free to intermix in the horizontal plane without the loss of alignment.

### 3. Transmission measurements

Transmission spectra of the films were then studied in the whole THz range (0.1–20 THz) using a combination of Terahertz Time-Domain Spectroscopy (THz-TDS) and Fourier-Transform Infra-Red spectroscopy (FTIR). In Fig. 3(a) we present the FTIR transmission spectra of a metamaterial film containing ordered metal wires (in the range 0.1–20 THz) for two different polarization of the incoming light. Grey color corresponds to polarization of the electric field parallel to the wires, whereas black color corresponds to polarization perpendicular to the wires. As seen from Fig.3 a) metamaterial films are strongly polarization sensitive for frequencies up to 5 THz, while polarization sensitivity persists up to 10 THz. For comparison, in Fig. 3(b) we present transmission spectra through a metamaterial film containing $As_2Se_3$ semiconductor glass wires, where no polarization sensitivity is observed.

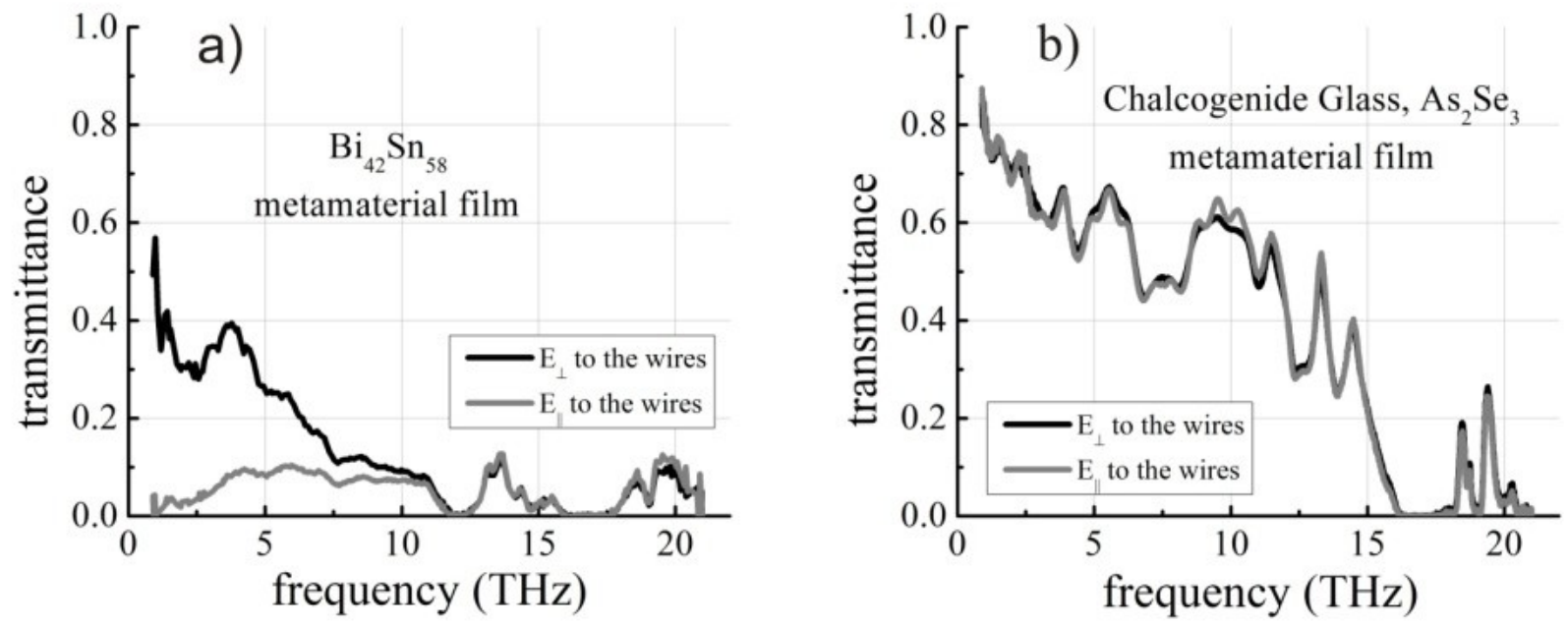

Fig. 3. FTIR transmission spectra (0.1–20 THz) of a metamaterial film containing ordered (a) metal and (b) semiconductor wires. Strong polarization dependence of transmission spectrum is observed for metallic wire arrays.

Polarization dependence of THz light transmission through metamaterial samples is clearly seen in Fig. 3(a); however, the lack of phase information makes it difficult to retrieve the corresponding effective refractive indices for the two polarizations. As shown in Fig. 3(a), light polarized parallel to the metal wires is blocked, while light polarized perpendicular to the wires is largely transmitted. The wire-grid polarizer consists of an array of slender wires arranged parallel to one another. The metal wires provide high conductivity for the electric fields parallel to the wires. Such fields produce electric currents in the wires, like the microwave dipole receiver antenna. The energy of the fields is converted into energy of currents in the wires. Such currents are then partially converted to heat because of the small but significant electrical resistance of the wires, and partially irradiated back. The physical response of the wire grid is, thus, similar to that of a thin metal sheet. As a result, the incident wave is mostly absorbed or reflected back and only weakly transmitted in the forward direction. Because of the nonconducting spaces between the wires, no current can flow perpendicular to them. So electric fields perpendicular to the wires produce virtually no currents and lose little energy, thus, there is considerable transmission of the incident wave. In this respect, the wire grid behaves as a dielectric rather than a metal sheet.

Alternatively, for the parallel polarization of incoming wave, response of a wire-grid medium can be modeled as that of a metal with an effective plasma frequency defined solely by the geometrical parameters of the medium such as wire size and inter-wire separation. Thus, at lower frequencies below the effective plasma frequency of a wire medium, electromagnetic waves are effectively blocked by such a material. However, at frequencies above plasma frequency polarization sensitivity of the wire grid medium is greatly reduced as both polarizations can go through the medium. At even higher frequencies when the inter-wire spacing and individual wire size (which are comparable in our samples) become comparable to the wavelength of light the metamaterial approximation breaks down. In our samples we observe that polarization sensitive transmission through the wire-medium is lost completely at frequencies higher than 10 THz.

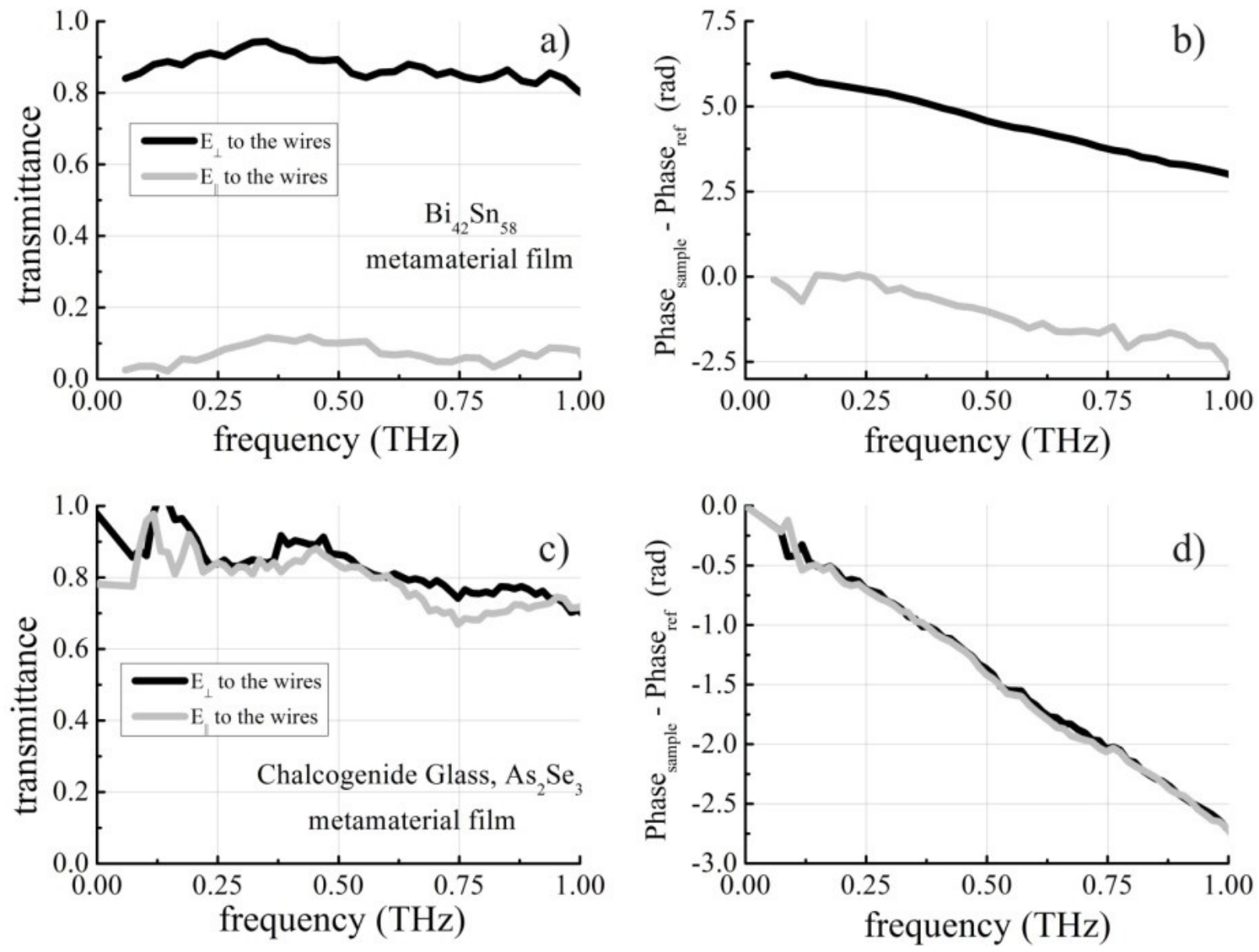

Fig. 4. Transmission spectra (a,c) and phase difference (b,d) of THz light through metamaterial film containing (a,b) ordered metallic wires, (c,d) ordered semiconductor wires.

In order to obtain complex permittivity of the samples, complex transmission spectra (amplitude (Fig.4 a,c) and phase(Fig. 4 b,d)) through metamaterial films were measured with a THz-TDS setup. The THz-TDS setup uses a frequency-doubled femtosecond fiber laser (MenloSystems C-fiber laser) as a pump source. Identical GaAs dipole antennae were used as THz source and detector yielding a spectrum ranging from 0.1 to 3 THz. The FWHM spot size of the THz beam at the focal point of the parabolic mirrors is roughly 3 mm. For convenient handling of samples, the metamaterial films were cut into 1cm x 1cm pieces and placed at the focal point of the parabolic mirrors. For each film sample, two transmission spectra were measured corresponding to the parallel and perpendicular polarizations of the THz beam with respect to the wire direction. To ensure constant input illumination conditions during rotation of samples, we used a rotation mount (RSP1 from Thorlabs). It utilizes two precision bearings for smooth, backlash-free rotation, and a small iris was inserted in the center of the rotation mount. The metamaterial film was then placed over the iris. Normalized transmittance through the metamaterial film was calculated by dividing the transmission data with, and without, the metamaterial film.

## 4. Interpretation of the experimental data

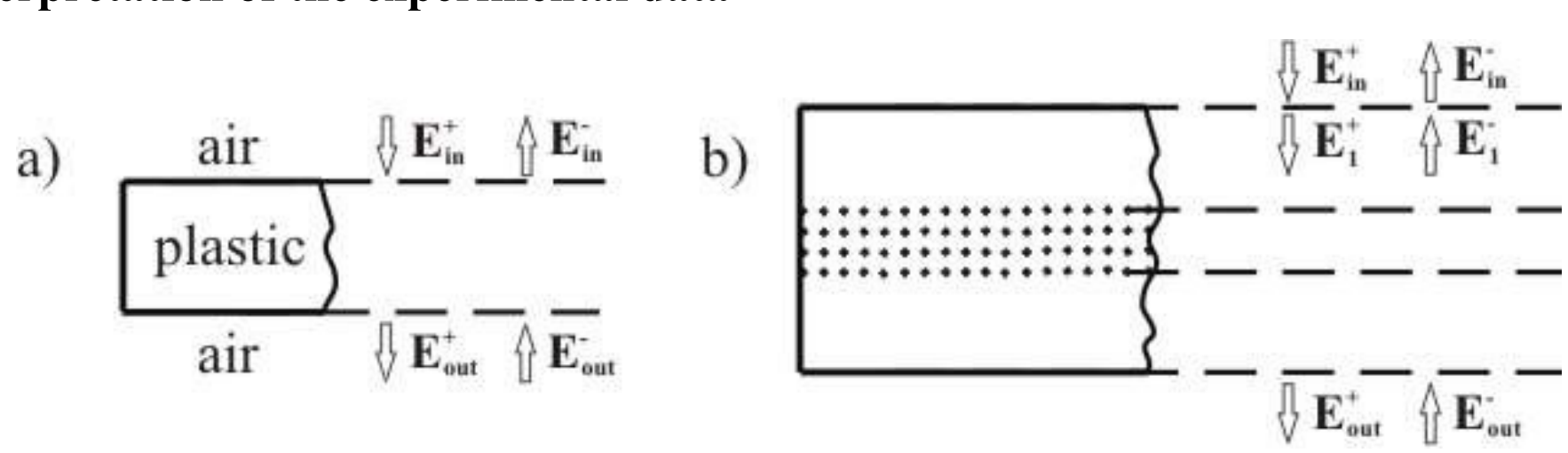

Fig. 5. Schematic of (a) one layr of plastic and (b) a metamaterial film modeled as a three layer system. The subscripts indicate the layer number, while the + and the – signs distinguish incoming an outgoing waves, respectively.

In order to infer the optical parameters of metamaterial layers, one has to assume a particular geometry of a composite film, as well as optical parameters of the constituent materials. As mentioned previously, the fabricated films can be viewed as a three-layer system (Fig.2 a, b), where a wire metamaterial layer is sandwiched between the two polymer layers (either PC or PSU for metal or chalcogenide wire films, respectively). Therefore, we first characterize refractive indexes and absorption losses of the pure PC and PSU polymers. We then use this data to fit the parameters of a metamaterial film.

*4.1. Procedure for retrieving the refractive index and losses of pure polymers*

Characterization of refractive indices of pure plastics is done with THz-TDS setup using thick polymer slabs with parallel interfaces. The samples were prepared by pressing the PC or PSU tubes into thick films at temperatures similar to those used during the fabrication of metamaterial films. Transmission spectra and phase difference of THz light through a plastic slabs is shown at Fig.6 a),b).

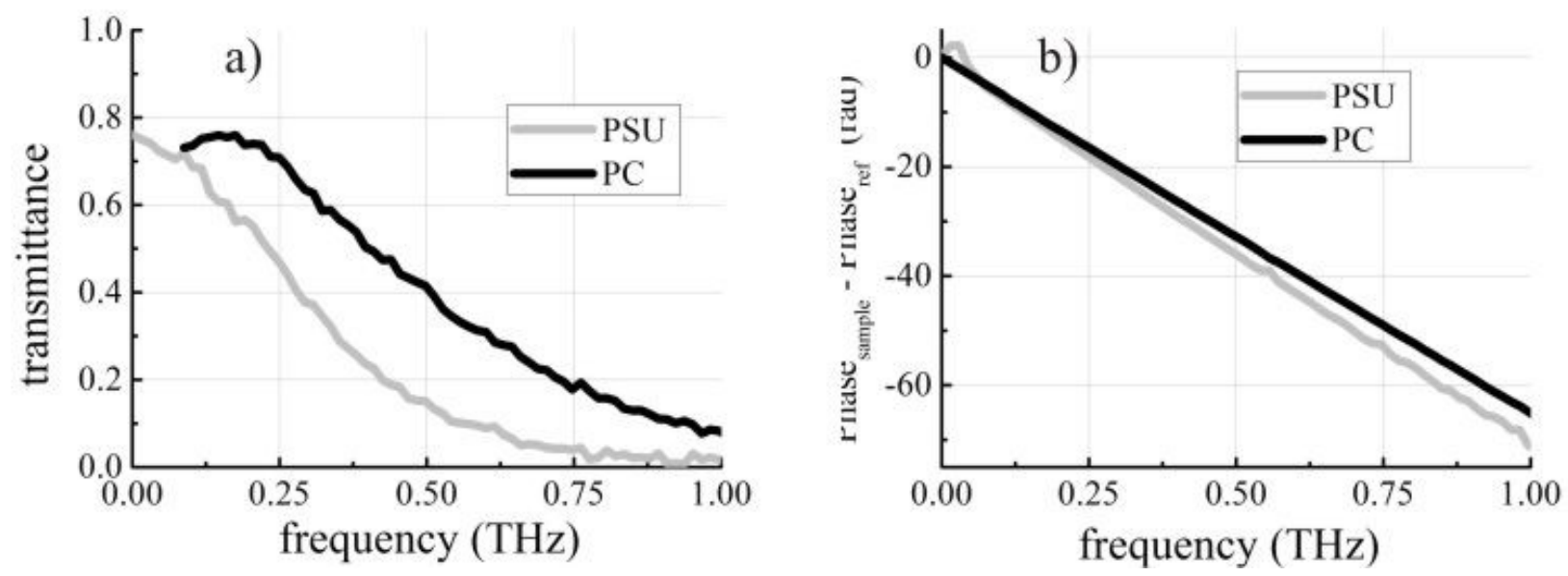

Fig. 6. Transmission spectra (a) and phase difference (b) of THz light through a plastic slabs.

We retrieve the refractive index and absorption losses of pure plastics by fitting the predictions of a transfer matrix model to the experimental transmission data [20–22]. Particularly, using the + and – superscripts to denote forward and backward traveling waves, the field amplitudes before and after a plastic slab are:

$$\begin{pmatrix} E_{in}^{+} \\ E_{in}^{-} \end{pmatrix} = \prod_{i=1}^{2} M_{i-1,1} \begin{pmatrix} E_{out}^{+} \\ 0 \end{pmatrix} = \begin{bmatrix} T_{11}^{m} & T_{12}^{m} \\ T_{21}^{m} & T_{22}^{m} \end{bmatrix} \begin{pmatrix} E_{out}^{+} \\ 0 \end{pmatrix} \quad (1)$$

The complex field transmission coefficient through a single layer is then given by

$$t = \frac{1}{T_{11}^{m}} = \frac{t_{a,p} t_{p,a} \exp(-i\, n_p \omega d_p / c)}{1 + r_{a,p} r_{p,a} \exp(-2i\, n_p \omega d_p / c)} \quad (2)$$

where $r_{a,p}, r_{p,a}$ and $t_{a,p}, t_{p,a}$ are the complex reflection and transmission Fresnel coefficients of the interface between air (subscript *a*) and plastic (subscript *p*), as well as between plastic and air respectively. These coefficients have the same form for both *s* and *p* polarizations due to normal angle of radiation incidence. Denoting $d_p$ to be the plastic layer thickness, and $n_p$ to be the refractive index of a plastic layer, we can write:

$$t_{a,p} = \frac{2n_a}{n_a + n_p} \quad (3)$$

$$r_{a,p} = \frac{n_a - n_p}{n_a + n_p} \quad (4)$$

For the calculation of refractive index and absorption losses of pure plastics we can assume that slabs are thick enough, so that multiple reflections within a sample (Fabry-Pérot reflections) can be neglected. In this case we can assume that the denominator of Eq. (2) simply equals 1, which allows us to write:

$$t = T(\omega)e^{i\phi} = \frac{4n^{real}(\omega)}{(n^{real}(\omega)+1)^2} \exp\left(-i\frac{\omega d_p}{c}\left(n^{real}-1\right)\right)\exp\left(-\frac{1}{2}\alpha d_p\right) \quad (5)$$

The transmittance $T(\omega)$ and phase shift $\varphi$ at each frequency are obtained simultaneously by the THz-TDS measurements. Eq. (5) easily separates into two equations respectively for the real and imaginary parts of the refractive index. For $n^{real}$ and $\alpha$ we get:

$$n^{real}(\omega) = 1 + \frac{c\phi}{\omega d_p} \quad (6)$$

$$\alpha = 2n^{im}\omega / c = \frac{2}{d_p}\ln\left(\frac{(n^{real}(\omega)+1)^2 T(\omega)}{4n^{real}(\omega)}\right) \quad (7)$$

It is important to realize that imaginary part of the refractive index of common plastics in THz frequency range can be several percent from the value of the real part of the refractive index. Therefore, to guarantee the third digit accuracy in the value of the real part of the extracted refractive index, Eq. (6,7) may no longer be valid, and the use of the complete formula Eq. (5) may be necessary. Experimentally the thicknesses of the two polymer slabs were 5 mm. In fact, the samples should be thick enough so that multiple Fabry-Pérot reflections are suppresses via absorption of the reflected waves on the length scale of a sample width. On the other hand, the sample should not be too thick, otherwise absorption losses even after one passage would be too large to make a reliable measurement of a transmitted field. The value of the fitted refractive index of PC is almost constant across the whole THz range and it equals to $n = 1.64$, which is in good agreement with prior reports [25]. Absorption loss of plastics are a rapidly increasing function of frequency which can be fitted as PC: $\alpha_{pc}[cm^{-1}] = 3.77f + 6.65f^2$, PSU: $\alpha_{psu}[cm^{-1}] = 5.03f + 20.25f^2$,where $f$ is frequency in [THz]. In this work, for the first time to our best knowledge, we also report the THz refractive index of PolySulphone $n = 1.74$, which is also effectively constant across the whole THz range. Fig.7 details frequency dependence of the PC and PSU refractive indices (a) and absorption losses (b) obtained by fitting the experimental transmission data.

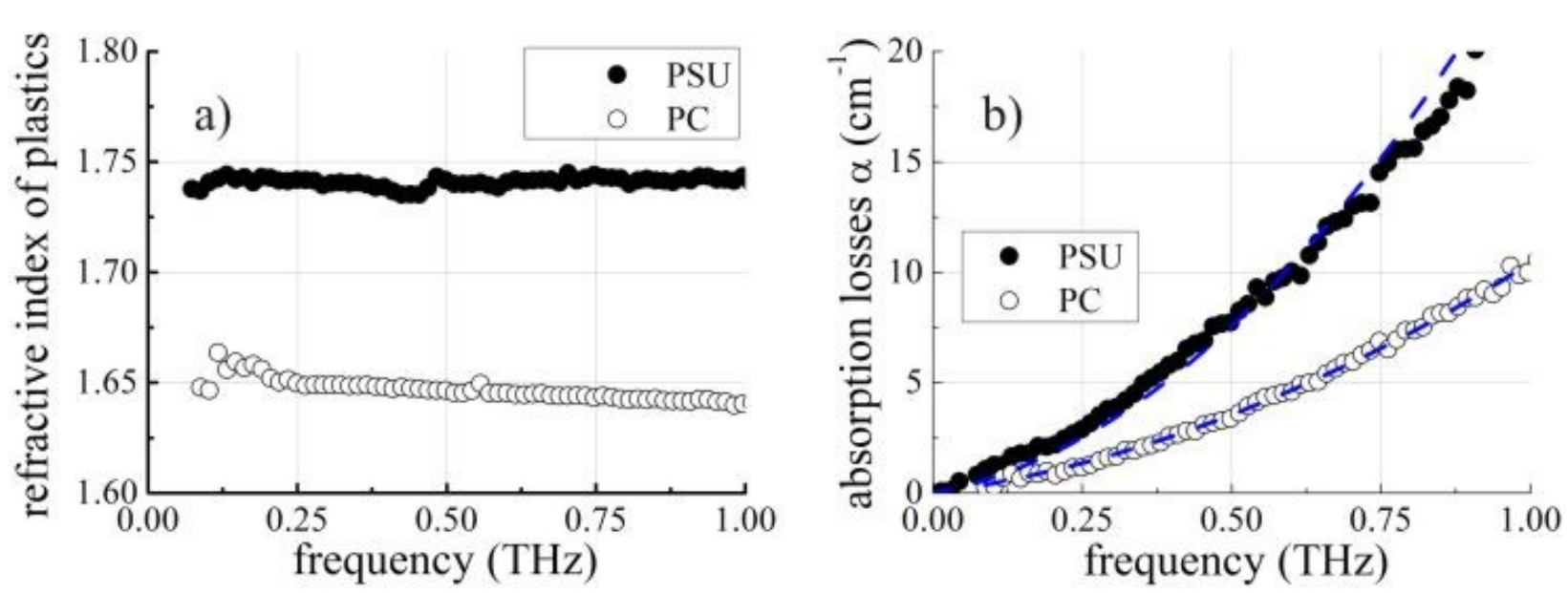

Fig. 7. a) real part of the refractive index and b) absorption losses of pure plastics PC, PSU.

*4.2. Procedure for retrieving the refractive index and permittivity of metamaterial layers*

As mentioned earlier, we model the experimental films as a three-layer system (Fig.5 b)) where a wire metamaterial layer is sandwiched between two polymer layers (either PC or PSU for metal or chalcogenide-based wire arrays). We start with the metal wire-based metamaterial films. The total thickness of such films is 150 µm. Thickness of a metamaterial layer alone is 35±5 µm. The layer comprises metal wires of 1.3±0.5 µm average diameter.

The complex transmission of a 3-layer system is modeled using a transfer matrix theory, assuming that a multilayer is surrounded by air. The field amplitudes $E_{in}$ , $E_{aut}$ in the first and last semi-infinite air regions respectively are related via the product of four 2×2 transfer matrices on each interface:

$$\begin{pmatrix} E_{in}^{+} \\ E_{in}^{-} \end{pmatrix} = \prod_{i=1}^{4} M_{i-1,1} \begin{pmatrix} E_{out}^{+} \\ 0 \end{pmatrix} = D_0^{-1} \left[ \prod_{i=1}^{3} D_i P_i D_i^{-1} \right] D_4 \begin{pmatrix} E_{out}^{+} \\ 0 \end{pmatrix} = \begin{bmatrix} T_{11}^{m} & T_{12}^{m} \\ T_{21}^{m} & T_{22}^{m} \end{bmatrix} \begin{pmatrix} E_{out}^{+} \\ 0 \end{pmatrix} \tag{8}$$

Each side of an interface is represented by the corresponding transmission matrix $D_i$, whereas the propagation inside the bulk material of each layer is represented by its propagation matrix $P_i$

$$D_i = \frac{1}{t_{i-1,i}} \begin{bmatrix} 1 & r_{i-1,i} \\ r_{i-1,i} & 1 \end{bmatrix} \tag{9}$$

$$P_i = \begin{bmatrix} \exp(i\frac{n_i \omega d_i}{c}) & 0 \\ 0 & \exp(-i\frac{n_i \omega d_i}{c}) \end{bmatrix} \tag{10}$$

where $r_{i-1,i}$ and $t_{i-1,i}$ are the complex reflection and transmission Fresnel coefficients:

$$t_{i-1,i} = \frac{2n_{i-1}}{n_{i-1} + n_i} \tag{11}$$

$$r_{i-1,i} = \frac{n_{i-1} - n_i}{n_{i-1} + n_i} \tag{12}$$

of the interface, $d_i$ is the thickness of *i* layer, and $n_i$ is the complex refractive index. In our case we made only transmission measurements, thus we only have amplitude and phase information about the forward traveling waves $E_{in}^{+}$ and $E_{out}^{+}$. The complex transmission coefficient of the multilayer is given in terms of the system transfer matrix elements $T_{ij}$ as $T_{theor.} = 1/T_{11}^{m}$ . The THz-TDS setup allows to obtain information about amplitude $E_{out}^{+}$ and phase $\phi$ of the signal transmitted through the sample for all THz frequencies in a single measurement. The measured complex transmission coefficient is obtained from the ratio of the sample and reference fields,

$$T_{measured} = T(\omega)e^{i\phi} = E_{out}^{+}(\omega) \big/ E_{ref}^{+}(\omega) e^{i(\phi-\phi_{ref})} \tag{13}$$

The reference field amplitude $E_{ref}^{+}$ and phase $\phi_{ref}$ are obtained by measuring the signal without sample._Assuming that multilayer geometry and refractive index of a host material are known, complex refractive index of the metamaterial layer is found by taking the difference between the measured and theoretical transmission coefficients to be zero, and then finding the roots of a resultant equation:

$$T_{measured} - T_{theory}\left(\mathrm{Re}\left(n_{meta}\right), \mathrm{Im}\left(n_{meta}\right)\right) = 0 \tag{14}$$

Note that equation (14) is effectively a system of two purely real equations for the real and imaginary parts of the complex transmission coefficient. This system of equations has two unknowns, which are the real and imaginary parts of the metamaterial refractive index. Therefore, equation (14) is well defined.

The theoretical transmission coefficient of the multilayer depends on the knowledge of the exact thicknesses of the layers and their complex refractive indices. While the complex refractive indices of the host polymers, as well as the total thicknesses of a multilayer have been accurately measured, the thickness of a metamaterial layer must be assumed to be an adjustable parameter since its accurate estimation from the micrographs is not possible due to ambiguity in the definition of a boundary between metamaterial and a plastic cover. As it will become clearer in the following section, the exact choice for the value of the thickness of a metamaterial layer can only be made if an additional optimization criterion is established. Without a supplemental restriction on Eq. (14), this equation can be resolved for any given value of the metamaterial layer thickness. To solve numerically Eq. (14) we use a nonlinear Newton method for the real and imaginary parts of the refractive index at each frequency point. Another subtlety in solving Eq. (14) comes from the possibility of multiple solutions of this equation. For the cases with more than one solution, unphysical solutions are eliminated by verifying their compatibility with the Kramers–Kronig relations. Finally, the effective permittivity is calculated from the extracted complex refractive index of the metamaterial in order to facilitate the interpretation of the metamaterial behaviour and the analysis via the effective medium theory.

### *4.3. Extracted refractive index and permittivity of the films*

Fig. 8 presents the complex refractive index and complex permittivity of the metal wire metamaterial films, calculated using the transmission spectra of Fig.4.a) and the procedure described in the previous section. A clear difference in the reconstructed optical properties of a metamaterial layer can be seen for the two polarizations of the incident light. Not surprisingly, we find that a wire-grid structure behaves like a metal ($\mathrm{Re}(\varepsilon) \prec 0$ and $\mathrm{Im}(\varepsilon) \succ 0$) for the electromagnetic waves polarized parallel to the wires. Alternative interpretation for the operational principle of the wire grid polarisers is in terms of the effective plasma frequency of a metamaterial. This was proposed by Pendry et al. [26] to explain wave propagation through a 2D array of wires. He demonstrated that such an array behaves like a Drude metal with effective plasma frequency which is only a function of the geometrical parameters of a wire array $\omega_p^2 = 2\pi c_0^2 \big/ \left(a^2 \ln(2a/d)\right)$, where $c_0$ – the vacuum light speed. Several more complicated but precise derivations of the effective plasma frequency for the wire medium have been developed [27] and a comparison between the models has been conducted in [28]. Since the effective plasma frequency in a wire medium can be tuned by adjusting the medium's geometrical parameters, the spectral region for the desired permittivity values can be engineered to occur at practically any frequency. That formula for plasma frequency of the wire-grid medium is used to describe highly ordered structures, disordering yields to shifting of it to the lower frequencies [29].

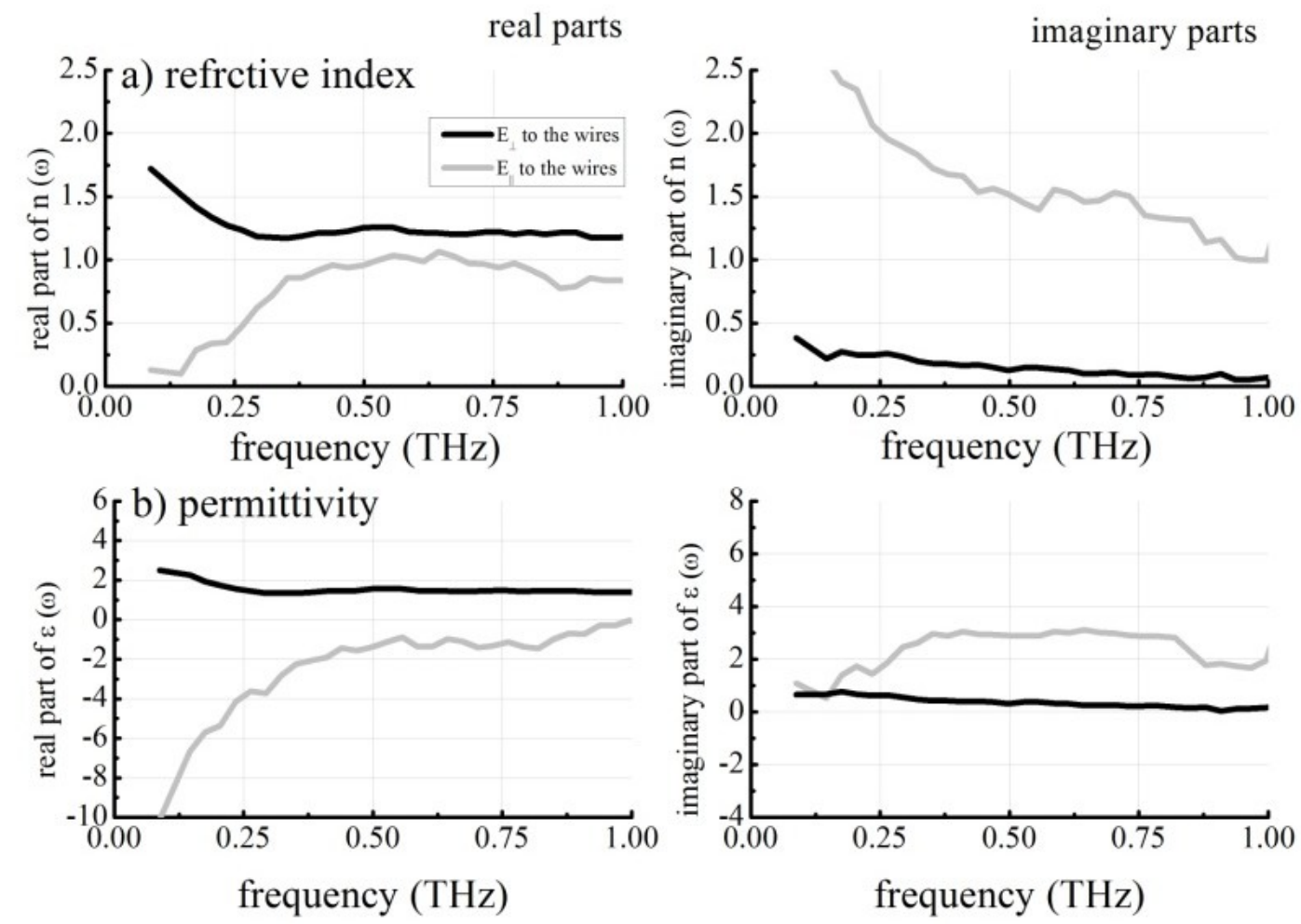

Fig. 8. Extracted (a) refractive index, and (b) permittivity of a metamaterial layer containing metal wires.

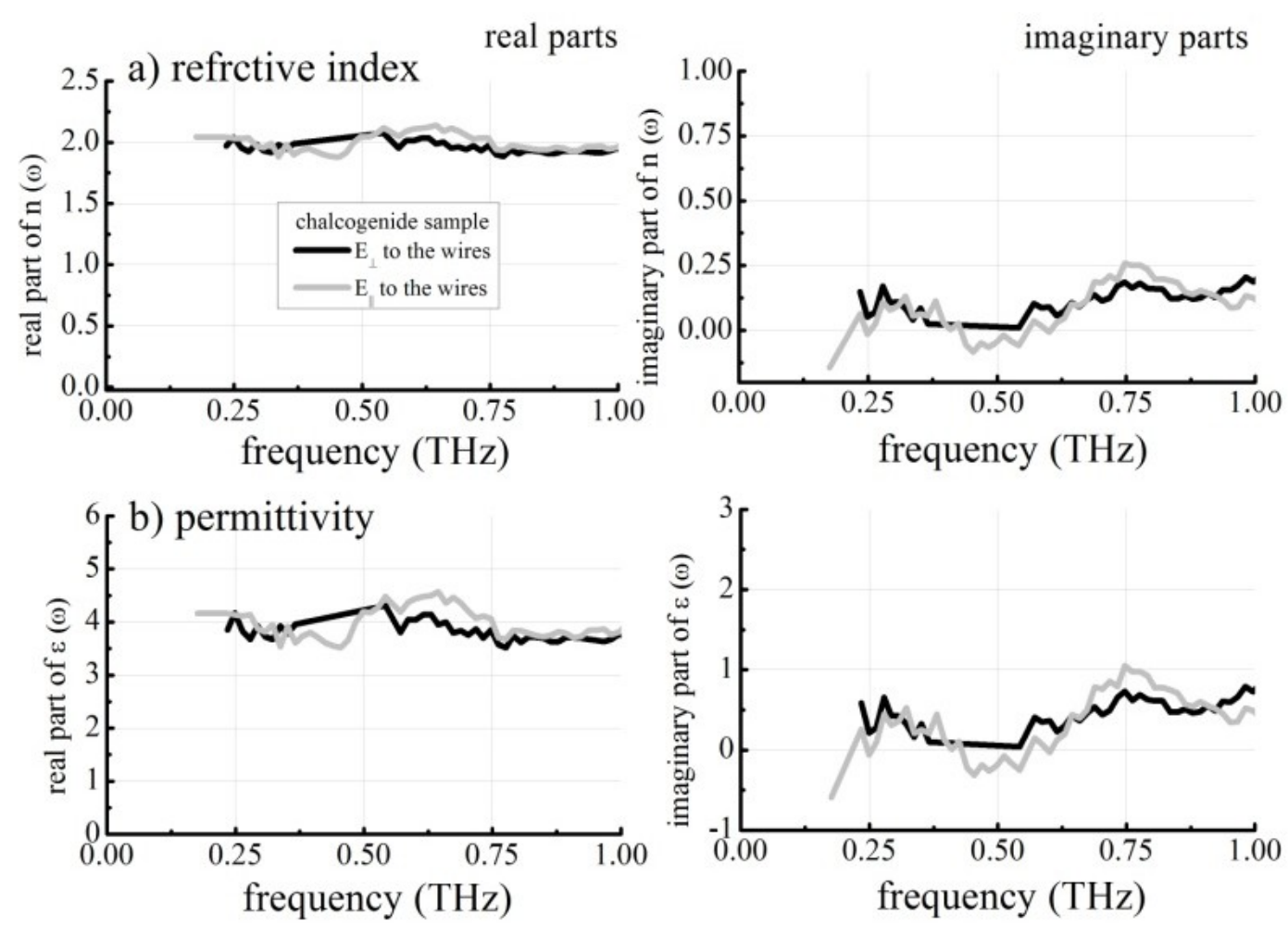

Fig. 9. Extracted (a) refractive index, and (b) permittivity of a metamaterial layer containing chalcogenide glass wires

For comparison, we considered the refractive index and permittivity of the metamaterial layer within the metamaterial film containing chalcogenide glass wires. The values calculated from the transmission spectra of Fig.4.b) are shown in Fig.9. Due to the low conductivity of the chalcogenide wires, the wire-array does not act as an effective metal and there is no difference between the polarizations parallel or perpendicular to the wires. In this case, the chalcogenide glass inclusions merely form a dielectric composite material having a significantly higher refractive index compared to the pure PSU plastic host.

*4.3. Effective medium theory*

In the previous subsection we employed a transfer matrix theory to extract effective refractive indices of the metamaterial layer from the optical transmission data through a three layer composite film. Using the effective medium theory [3,4,30] one should also be able to obtain the optical parameters of a metamaterial layer from the optical parameters of the constituent plastic cladding and metal/dielectric inclusions, and from the geometry and density of the inclusions creating a composite material. Inversely, knowing the polarization dependent optical transmission properties of the metamaterial layer one should be able to extract the optical properties of the constituent materials (metals/dielectrics). In this subsection we detail the use of an effective medium theory to extract the effective refractive index (see Fig. 8) of the wire material. In fact, we propose a self-consistent algorithm that for every one of the two polarizations studied in this paper predicts the spectral dependence of the wire material dielectric constant, and then makes sure that the difference between the two predictions is minimized through adjustment of various geometrical parameters of a metamaterial layer.

There are many effective medium theories; all of them require an a priori knowledge of the permittivities of the dielectric host and metal inclusions. In our work, the material for the metal-wire arrays we use is a $Bi_{42\%}Sn_{58\%}$ eutectic solder alloy. In the existing literature we did not find the dielectric data for this alloy in the THz range. In this subsection we demonstrate how to retrieve the optical data for this metallic alloy in a self-consistent manner from the two independent measurements using parallel and perpendicular polarizations with respect to the wire direction. Our goal is to achieve consistency between these two estimates by using an appropriate choice of the effective medium parameters.

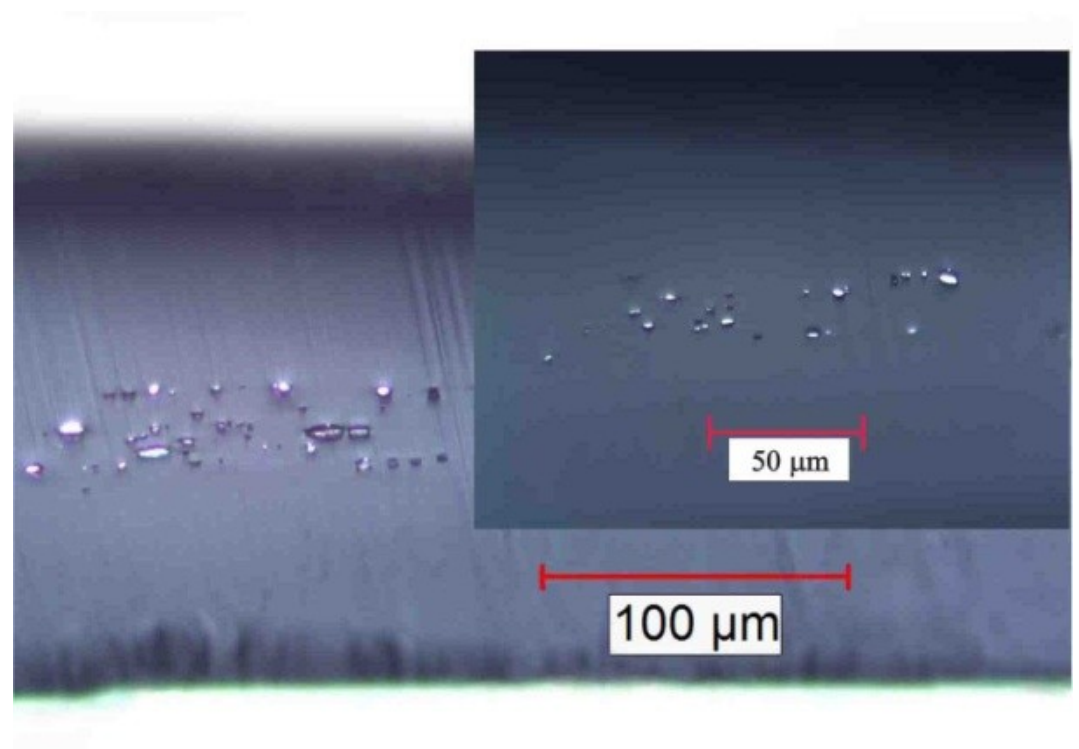

Fig. 10. Optical micrograph of the metamaterial film containing metal microwire array

A typical effective medium theory assumes that inclusions (spheres, discs, rods) are randomly dispersed within a host matrix, and that these inclusions are much smaller than the wavelength of light inside the host material (<0.1– 0.2 λ), such that the material can be treated like an effective dielectric medium. In our case, the incident light of frequency 1 THz has a corresponding wavelength of 300 μm, which is almost 100 times lager that the average wire diameter of 1.3 μm. Among many effective medium theories, the most common is the Maxwell-Garnett model [31-33], which uses as a fundamental parameter a volume fraction (filling factor) $f = V_{particles} / V_{background}$ of the particles embedded in the uniform background material. In the case of wire arrays, $V_{background}$ is proportional to the active area of the background material, and $V_{particles}$ is proportional to the total area of the metallic wires. For the polarization parallel to the wires, the effective permittivity of the composite in Maxwell-Garnett approach is then given by:

$$\left( \frac{\varepsilon_{eff}(\omega) - \varepsilon_{plastic}}{\varepsilon_{eff}(\omega) + k\varepsilon_{plastic}} \right) - f \left( \frac{\varepsilon_m - \varepsilon_{plastic}}{\varepsilon_m + k\varepsilon_{plastic}} \right) = 0 \quad (15)$$

where $\varepsilon_m$ is the dielectric constant of the inclusions, $\varepsilon_{plastic}$ is the dielectric constant of the host material, and $k$ is a screening parameter. The value for $k$ is 1 if the inclusions are an array of cylinders, whose axes are collinear with the polarization of the incident radiation. For polarization perpendicular to the wires the effective permittivity of the composite is:

$$\varepsilon_{eff}(\omega) = f\varepsilon_m + (1-f)\varepsilon_{plastic} \tag{16}$$

The filling factor 0.15 of the metamaterial layer was estimated from optical micrograph of the metamaterial film (Fig.10). Note that the metamaterial layer is sandwiched between the two polymer layers, and we took filling factor as filling factor of the metamaterial layer only, and not the whole sample. For a fixed thickness of a metamaterial layer, to find the optimal value of the filling factor, we first eliminate the dielectric constant of the metal for polarization parallel to the wires from Eq.(15):

$$\varepsilon_m^{\parallel}(\omega) = \frac{\varepsilon_{plastic}(\omega)(\varepsilon_{eff}^{\parallel}(\omega)(1+f) + (f-1)\varepsilon_{plastic}(\omega))}{\varepsilon_{plastic}(\omega)(1+f) + \varepsilon_{eff}^{\parallel}(\omega)(1+f)} \tag{17}$$

For polarization perpendicular to the wires the dielectric constant of the metal is:

$$\varepsilon_m^{\perp}(\omega) = \frac{\varepsilon_{eff}^{\perp}(\omega) - (1-f)\varepsilon_{plastic}(\omega)}{f}. \tag{18}$$

We then form a weighting function

$$Q(f) = \int_{\omega_{\min}}^{\omega_{\max}} d\omega \left|\varepsilon_m^{\perp}(\omega) - \varepsilon_m^{\parallel}(\omega)\right|^2 \tag{19}$$

that we minimize with respect to the filling fraction. Ideally, if the metamaterial layer thickness and filling fractions are chosen correctly one expects $\varepsilon_m^{\perp}(\omega) = \varepsilon_m^{\parallel}(\omega)$, $Q = 0$.

Practically, the weighting function is always larger than zero, and minimization of the weighting function has to be performed both with respect to the filling fraction and thickness of the metamaterial layer. Solution of this optimization problem gives the optimal values of the filling factor $f$ =0.23, and optimal value of the metamaterial layer thickness of $d_{meta}$=35μm, which is in the range of values found experimentally from the micrograph analysis. In Fig. 11 we present comparison of the real and imaginary parts of the metal alloy as calculated using Eq. (17), Eq. (18).

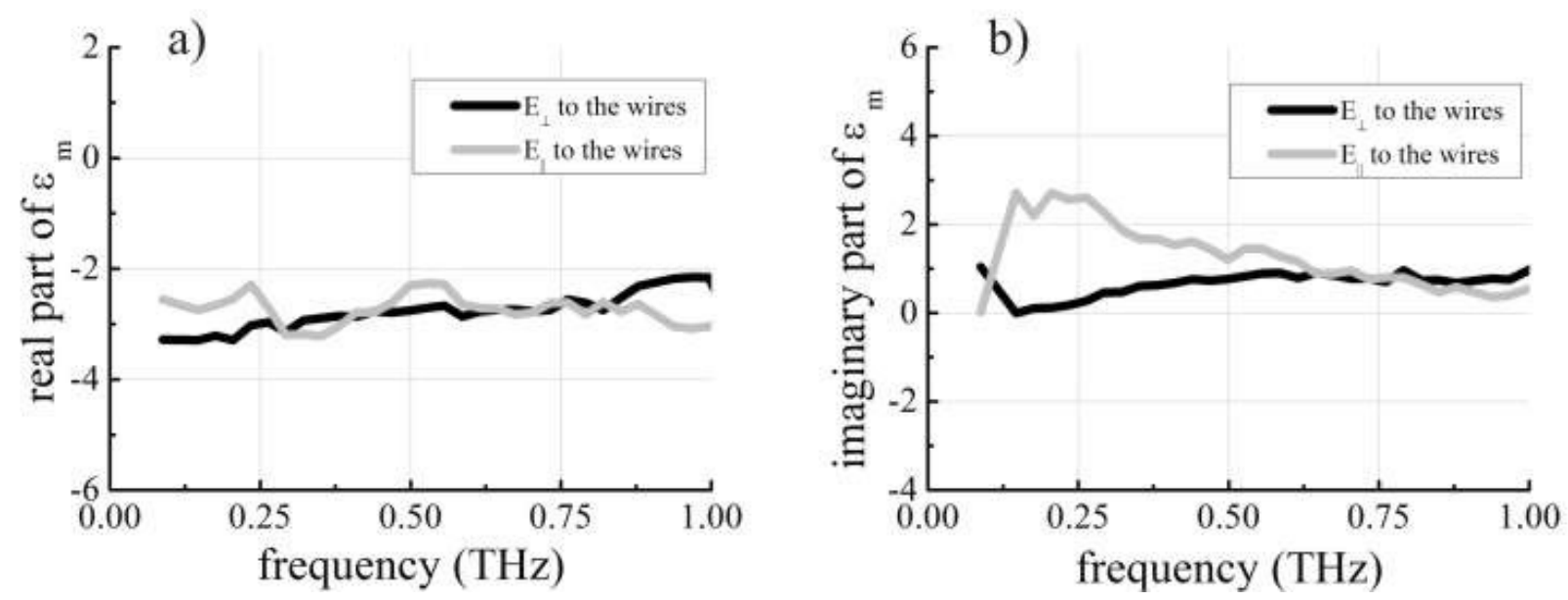

Fig. 11. The real (a) and imaginary parts (b) of the metal alloy as calculated for different polarizations

*4.4 Nanostructured inclusions*

We note in passing, that metallic wires made of multi-components alloys can exhibit an additional inherent nanostructure. Particularly upon solidification, the metals of $Bi_{42\%}Sn_{58\%}$ alloy have been found to separate into different pure metal phases [34]. To illustrate this,

Fig. 12 presents a scanning-electron-micrograph of eutectic alloy wires obtained after two subsequent drawings. These wires were extracted from the composite by dissolving the PC cladding. On the wire surfaces one clearly sees formation of nanogratings. Stoichiometric studies using TOF-SIMS measurements (Time-of-Flight Secondary Ion Mass Spectrometry) showed that the lamellar structure on the surface of the individual wires is made of pure Bi (Bismuth) and Sn (Tin) parts.

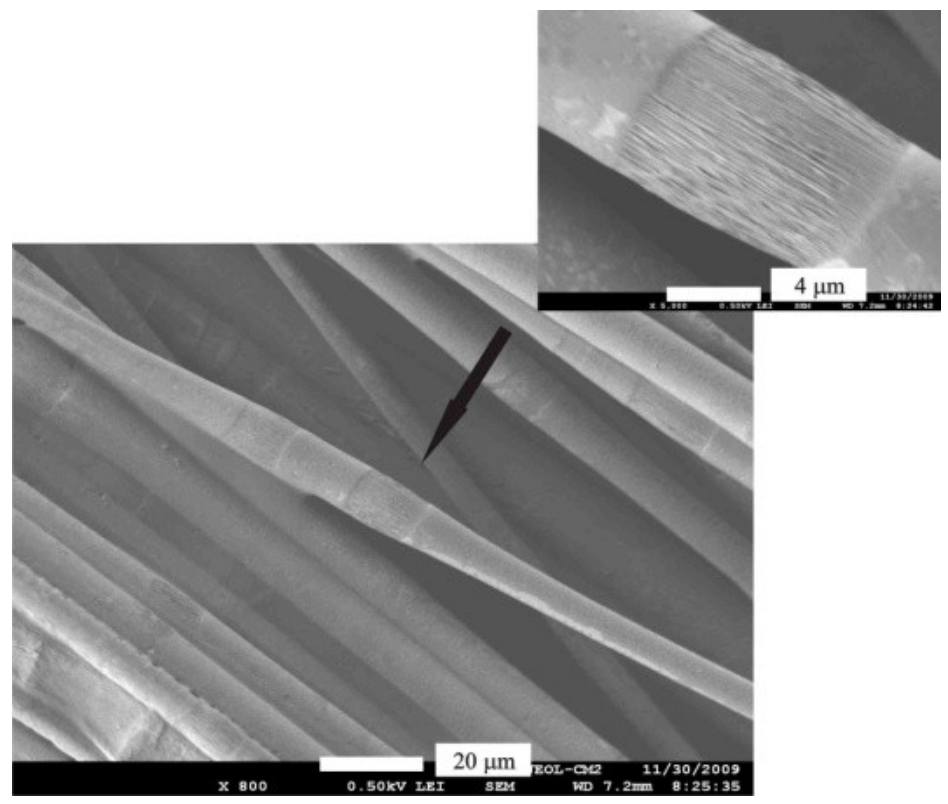

Fig. 12. Microwires after 2nd drawing featuring nanogrids. Microwires were obtained by dissolving PC polymer matrix. In the inset, shown is a pure Bi, Sn lamellar structure due to phase separation of an alloy.

## 5. Conclusion

Overall, this paper presents an in-depth study of the challenges and practical issues encountered when trying to extract from the THz optical transmission data both the material and geometrical parameters of the deeply subwavelength wire arrays making a metamaterial film. Particularly, we detail a novel fabrication method for the large area THz metamaterial films using a two step fabrication process. First, we draw microstructured fibers containing ordered arrays of micro- or nano- wires made from either metal or dielectric materials. We then arrange such fibers parallel to each other and compactify them into a composite film by using a hot-press technique. Characterization of the THz transmission through metal wire composite films demonstrates strong polarization sensitivity of their optical properties, promising application of such films as polarizers or high pass filters for THz radiation. Furthermore, composite films containing high refractive-index dielectric wires in place of metallic wires showed no polarization sensitivity, while the effective refractive index of such metamaterial films could be adjusted in a broad range. Using the transfer matrix theory, we then demonstrate how to retrieve the polarization dependent complex refractive index and complex effective permittivity of the metamaterial films from the THz optical transmission data. Finally, we detail the selfconsistent algorithm for retrieving the optical properties of the metal alloy used in the fabrication of the metamaterial layers by using an effective medium approximation.